\documentclass[10pt,preprintnumbers,showpacs,amsmath,amssymb,floatfix,twocolumn,prl]{revtex4}
\usepackage{graphicx}% Include figure files
\usepackage{dcolumn}% Align table columns on decimal point
\usepackage{bm}% bold maths
\usepackage{longtable}
\usepackage{graphics}
\usepackage{amssymb}
\usepackage{amsmath}
\usepackage{xspace}
\usepackage{epsfig}
\newcommand {\Brn}{$\kappa$-(BEDT-TTF)$_2$Cu[N(CN)$_2$]Br}

\begin{document}
\title{Electron-phonon coupling as an order-one problem}
\author{B. J. Powell}
\email{powell@physics.uq.edu.au} \affiliation{Department of
Physics, University of Queensland, Brisbane, Queensland 4072,
Australia}
\author{Mark R. Pederson}
\affiliation{Center for Computational Materials Science, Code
6392, Naval Research Laboratory, Washington DC 20375-5345}
\author{Tunna Baruah}
\affiliation{ Center for Computational Materials Science, Code
6392, Naval Research Laboratory, Washington DC 20375-5345}
\affiliation{Department of Physics, SUNY Stony Brook, NY, 11794-3800}

%\pacs{87.15.-v, 82.35.Cd, 87.64.Je}

\begin{abstract}
The coupling between electrons and phonons plays important roles in
physics, chemistry and biology. However, the accurate calculation of
the electron-phonon coupling constants is computationally expensive
as it involves solving the Schr\"odinger equation for ${\cal O}(N)$
nuclear configurations, where $N$ is the number of nuclei. Herein we
show that by considering the forces on the nuclei caused by the addition or
subtraction of an
arbitrarily small electronic charge one may calculate
the electron-phonon coupling constants from ${\cal O}(1)$ solutions
of the Schr\"odinger equation. We show that Janak's theorem means
that this procedure is exact within the density functional
formalism. We demonstrate that the ${\cal O}(1)$ approach produces
numerically accurate results by calculating the electron-phonon
coupling constants for a series of molecules ranging in size from
H$_2$ to C$_{60}$. We use our approach to introduce a computationally
fast approximation for the adiabatic ionisation potentials and electron
affinities which is shown to be accurate for large molecules. We also
show that
our approach allows for the  calculation isotope effects in ${\cal
O}(0)$ time and  discuss the deuteration driven Mott transition in
\Brn.
\end{abstract}

\maketitle

Electrons in condensed matter feel two forces: a direct Coulomb
interaction between themselves and an Columbic interaction with the
atomic nuclei. As the vibrations of the nuclei are quantised the
interaction with the lattice takes the form of the electron-phonon
(or electron-vibron) interaction. Electron-phonon interactions plays
important roles across science from physics (electron-phonon
interactions can give rise to superconductivity \cite{Ziman}, spin
and charge density waves \cite{Gruner}, polaron formation
\cite{Mott&Alexandrov} and piezoelectricity \cite{Ziman}), through
chemistry (such as in electron transfer processes \cite{Marcus},
Jahn-Teller effects \cite{Bersuker}, spectroscopy \cite{Bersuker},
stereochemistry \cite{Bersuker}, activation of chemical reactions
\cite{Bersuker} and catalysis \cite{Bersuker}) to biology (for
example electron-phonon interactions play an important role in
photoprotection \cite{Meredith}, photosynthesis \cite{Reimers} and
vision \cite{rhodopsin}). It is therefore clear that one of the
central tasks for condensed matter theory and theoretical chemistry
is to accurately calculate electron-phonon coupling constants.
However, until now, this has been a computationally expensive task.

The expense of calculating the electron-phonon coupling constants
arises from the large number of phononic modes that need to be
considered. For example, a (non-linear) molecule with $N$ atoms has
$M=3N-6$ phononic modes. Within the Born-Oppenheimer approximation,
the standard, frozen phonon, approach to calculating the
electron-phonon coupling constants requires the electronic
eigenstates to be calculated for $M$ nuclear configurations. Below
we demonstrate how the number of calculations can be reduced from
$M$ to 1 without increasing the difficulty of the electronic
problem.

Let us begin by reviewing the basic problem of electron-phonon
coupling in the semiclassical limit. In this limit the phonons are
represented by harmonic oscillators, thus the Hamiltonian is
\begin{eqnarray}
{\cal H}={\cal H}_e^0+\sum_i\left(\frac{p_i^2}{2m_i}
+\frac12k_ix_i^2 +\tilde{g_i}x_in\right)
\end{eqnarray}
where $p_i$ is the momentum of the $i^{th}$ mode, $m_i$ is the
effective mass of the $i^{th}$ mode, $k_i$ is the spring constant of
the $i^{th}$ harmonic oscillator, $x_i$ is the displacement of the
$i^{th}$ mode from its equilibrium position, the $\tilde g_i$ are
the electron-phonon coupling constants, $n$ is the number of
electrons and ${\cal H}_e^0$ is the Hamiltonian of the electrons in
the absence of phonons. In what follows it will be useful to introduce
dimensionless normal coordinates for the phonons given by
$Q_i=x_i\sqrt{m_i\omega_i/\hbar}$, where $\omega_i=\sqrt{k_i/m_i}$
is the frequency of the phonon, dimensionless momenta,
$P_i=p_i\sqrt{m_i/\hbar\omega_i}$, and dimensionless electron-phonon
coupling constants, $g_i=\tilde g_i/\sqrt{2m_i\hbar\omega_i^3}$.
Thus, the Hamiltonian may be written as
%\begin{eqnarray}
%{\cal H}={\cal H}_e^0\frac{\hbar\omega}{2}(P^2+Q^2)
%+\sqrt{2}g\hbar\omega Qn
%\end{eqnarray}
%This is readily generalised to the case of many phonons by
%considering multiple harmonic oscillators, in which case the
%Hamiltonian becomes
\begin{eqnarray}
{\cal H}={\cal
H}_e^0+\sum_i\left[\frac{\hbar\omega_i}{2}(P_i^2+Q_i^2)
+\sqrt2g_i\hbar\omega_i Q_in \right].
\end{eqnarray}
We may quantise this Hamiltonian by introducing the Bosonic
operators $\hat a_{i}^{(\dagger)}$ which annihilate (create) a
phonon on the in the $i^{th}$ mode and the Fermionic operators $\hat
c_{\mu}^{(\dagger)}$ which annihilate (create) an electron in the
state $\mu$ such that $\hat{n}_\mu=\hat c_{\mu}^\dagger \hat
c_{\mu}$, $\hat Q_i=\frac1{\sqrt 2}(\hat a_{i}^\dagger+\hat a_{i})$
and $\hat P_i=\frac{i}{\sqrt 2}(\hat a_{i}^\dagger-\hat a_{i})$.
Thus the quantum Hamiltonian is
\begin{eqnarray}
\hat{\cal H}=\hat{\cal H}_e^0 + \sum_i\hbar\omega_i \left(\hat
a_{i}^\dagger \hat a_{i} + \frac12 \right) \notag \\ +\sum_{i\mu}
g_{i\mu}\hbar\omega_i\hat c_{\mu}^\dagger \hat c_{\mu} (\hat
a_{i}^\dagger+\hat a_{i}).
\end{eqnarray}

The coupling between the $i^{th}$ phononic mode and the $\mu^{th}$
electronic state is given by
\begin{eqnarray}
g_{i\mu}=\frac{1}{\sqrt{2}\hbar\omega_i}\frac{\partial\lambda_\mu}{\partial
Q_i}  \label{eqn:dedQ}
\end{eqnarray}
where $\omega_i$ is the frequency of the mode and $\lambda_\mu$ is
the energy of the $\mu^{th}$ electronic state. Therefore, to
calculate the electron-phonon coupling constants via the frozen
phonon method one begins by calculating the electronic structure and
the normal modes of the system; one then makes a small displacement
of the system along each normal coordinate, and thus calculates
${\partial\lambda_\mu}/{\partial Q_i}$. Therefore if there are $N$
normal modes $N$ such calculations must be performed. The frozen
phonon method can be speeded up by the use of density functional
perturbation theory (DFPT) \cite{DFPT}. However, DFPT is an ${\cal
O}(N)$ technique, albeit with a smaller coefficient than the frozen
phonon calculation. Below we present a new method to calculate
${\partial\lambda_\mu}/{\partial Q_i}$ in a single calculation. %We
%then use this method to calculate the electron-phonon coupling
%constants for C$_{60}$, tetrathiafulvalene (TTF),
%tetracyanoquinodimethane (TCNQ), and
%bis(ethylenedithio)tetrathiafulvalene (BEDT-TTF) and a number of
%smaller molecules and demonstrate that our method gives the same
%results as frozen phonon calculations. We also discuss the
%conceptual basis of the ${\cal O}(1)$ method and an approximation
%made useful by the speed of this method.

Janak's theorem  \cite{Janak} states that
\begin{eqnarray}
\lambda_\mu=\frac{\partial E}{\partial n_\mu} \label{eqn:Janak}
\end{eqnarray}
where $E$ is the total energy of the system and $n_\mu$ is the
electronic occupancy of the state $\mu$. It is important to realise
that, within the density functional formalism $n_\mu$ is not
required to be an integer \cite{Janak}. It follows from
(\ref{eqn:dedQ}) and (\ref{eqn:Janak}) that
\begin{eqnarray}
g_{i\mu} = -\frac{1}{\sqrt{2}\hbar\omega_i}\lim_{\delta
n\rightarrow0}\frac{1}{\delta n_\mu} \left(\frac{\partial
E}{\partial Q_i}\right)_{\delta n_\mu} \label{eqn:g}
\end{eqnarray}
where $(\partial E/\partial Q_i)_{x}$ indicates that the derivative
is taken after the charge is changed by $x$ relative to the charge
of the initially optimised geometry. We have used the fact that for
a (meta)stable geometry $(\partial E/\partial Q_i)_0=0$. Once the
electronic structure is solved in the equilibrium geometry of the
charge neutral system with a small change in the charge the forces
on the molecule can be calculated using the Hellman-Feynman theorem
and, as the dynamical matrix is already known from the calculation
of phonon spectrum, the electron-phonon coupling constants can
readily be calculated.

Conceptually it is useful to realise that in the frozen phonon and
DFPT approaches one considers the problem from the point of view of
the electrons. That is one calculates the electron-phonon coupling
constants by making a small perturbation to nuclei and considering
the resultant change on the electron. However, in the ${\cal O}(1)$
approach one asks the question what happens to the nuclei when one
puts a small additional charge in the system. In most cases one is
only interested in the coupling of the phonons to the electronic
states closest to the Fermi level, however, if one where interested
in couplings to other states extensions of Janak's theorem to other
electronic states allows for the determination of these parameters
straightforwardly within the ${\cal O}(1)$ method.

Thus  the main result of this letter is that using equation
(\ref{eqn:g}) the $g_i$ can be calculated by solving a single
nuclear geometry. In the remainder of this work we present some
benchmark calculations which show that equation (\ref{eqn:g})
reproduces the values of the electron-phonon coupling constants
calculated by the frozen phonon method for a number of small
molecules. To do this we have implemented our scheme for calculating
electron-phonon coupling constants in the Naval Research Laboratory
Molecular Orbital Library (NRLMOL)
\cite{NRLMOL1,NRLMOL2,NRLMOL3,NRLMOL4,NRLMOL5,NRLMOL6,NRLMOL7}.
Throughout we have used the Perdew-Burke-Ernzerhof (PBE) \cite{PBE}
exchange correlation functional.

\begin{figure}
\begin{center}
\epsfig{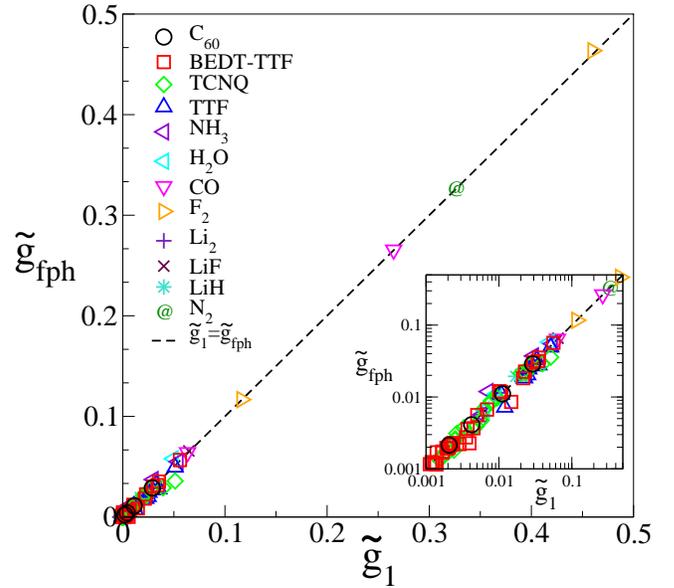}
\end{center}
\caption{Comparison of the electron-phonon coupling constants for a
variety of molecules calculated by the order-one method, $\tilde
g_1$, and the frozen phonon method $\tilde g_{fph}$ (both in
Ha/Bohr). It can be seen that the agreement between the frozen
phonon and order-one methods is excellent. In the inset we plot the
same data on a log-log scale which displays the excellent agrement
achieved even for small values of $\tilde g$. For each molecule we
have calculated electron-phonon interactions associated with the
HOMO and LUMO levels.} \label{fig1}
\end{figure}

In figure \ref{fig1} we plot the calculated electron-phonon
interactions for tetrathiafulvalene (TTF), tetracyanoquinodimethane
(TCNQ), bis(ethylenedithio)-tetrathiafulvalene (BEDT-TTF), C$_{60}$
and a number of diatomic and other small molecules calculated by the
frozen phonon method against the electron-phonon coupling constants
calculated by our method. It can be seen that the values calculated
by either method are the same within numerical noise.

Note that, as the normal coordinate is the eigenvector of the
dynamical matrix, the sign of $Q_i$ is not well defined. Therefore,
although $\partial E$ is negative definite, the sign of $g_{i\mu}$
is also not well defined. However, the product $g_{i\mu}\partial
Q_i$ is well defined and describes the distortion caused by the
addition of a charge. For example, for the stretching mode in a
simple diatomic molecule, the sign of the product $g_{i\mu}\partial
Q_i$ indicates whether the bond is lengthened or shortened when a
charge is added.

\begin{figure}
\begin{center}
\epsfig{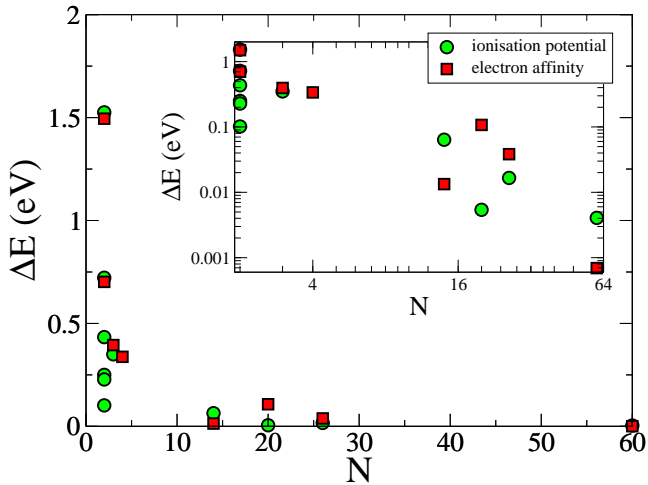}
\end{center}
\caption{The ionisation potentials and electron affinities of the
molecules considered in Fig. \ref{fig1}. We plot the difference in
energy, $\Delta E$ between the correct result (within the PBE
approximation) and the results from approximating the ionisation
potential (electron affinity) by estimating the Hubbard $U$ and
relaxation energy from the ${\cal O}(1)$ method. It can be seen that
the approximation improves dramatically as the number of atoms in
the molecule, $N$, increases. The inset show the same data on a
log-log scale, this suggests that the error in the approximation
decreases as $\Delta E\sim N^{-\alpha}$ with $\alpha\sim1.5-2$.}
\label{fig:IPEA}
\end{figure}

Because the calculation of electron-phonon coupling constants is now
computationally inexpensive, it is possible to use the
electron-phonon coupling to estimate other quantities. For example
the adiabatic ionisation energy (or electron affinity) of a
molecule, as opposed to the vertical ionisation energy (electron
affinity) may be extracted from the information already calculated.
To do this one determines the Hubbard $U$ (a second derivative) from
either the energy as a function of $\mp\delta n_\mu$ and $\mp
2\delta n_\mu$ or from the appropriate eigenvalue as a function of
$\mp\delta n_\mu$ As the electron-phonon coupling constants are
known the adiabatic ionisation energy (electron affinity) can be
calculated directly from Markus--Hush theory. In figure
\ref{fig:IPEA} we plot the difference between the adiabatic
ionisation energies (electron affinities)  calculated in this way
and the correct result (within the PBE framework) for the molecules
considered in Fig. \ref{fig1} as a function of the the number of the
atoms in the molecules. While for small molecules this approximation
is not accurate, the results for the larger molecules are excellent.
Clearly it is for larger molecules that this approximation is needed
as the computational power required to relax the geometry grows with
the size of the molecule.

It is important to stress that the calculation of the
electron-phonon coupling constants is, in fact, intrinsically an
${\cal O}(1)$ problem as the dynamical matrix and the forces
calculated on the geometries used to calculate the dynamical matrix
together with the eigenvalue changes determined during the course of
these calculations
contain all the information required to calculate the $g_{i\mu}$.
However, significant computational complexities must be overcome
to retrieve this information. Further, the calculated
electron-phonon coupling is rather sensitive to the size of the
displacement used in a frozen-phonon calculation. In general a simple criterion
for the size of the displacement, e.g., $x=\sqrt{2\Delta/k}$ where
$\Delta$ is a small energy, does not produce uniformly reliable results for all
possible phonon frequencies in
the frozen phonon calculations of the type discussed above. However,
we have found that the results for the ${\cal O}(1)$ method are
highly insensitive to the value of the `small' charge, $\delta
n_\mu$, used in the calculation.

Finally we note that the calculation of isotope effects is an ${\cal
O}(0)$ problem in our method, i.e. only trivial matrix manipulation
and no further solutions of the Schr\"odinger equation are required
to calculate the electron-phonon coupling constants as the isotopic
masses are varied. This contrasts to the frozen-phonon method where,
as the dynamical matrix and thus the normal modes change upon
isotopic substitution, ${\cal O}(N)$ additional calculations are
required.

A particularly interesting isotope effect is observed in the
superconductor $\kappa$-(BEDT-TTF)$_2$Cu[N(CN)$_2$]Br. When the
eight hydrogen atoms in the BEDT-TTF molecule are replaced by
deuterium the ground state is found to be a Mott insulator
\cite{Taniguchi}. In the crystal the highest occupied molecular
orbital (HOMO) of the BEDT-TTF molecule is doped with holes from the
anion layer, therefore one expects \cite{Mott&Alexandrov} that, in
the small polaron limit, on deuteration, the bandwidth will change
by a factor
\begin{eqnarray}
\frac{W_D}{W_H}=\frac{e^{-\sum_ig_{ihD}^2}}{e^{-\sum_ig_{ihH}^2
}}=0.97
\end{eqnarray}
where $W_D$ ($W_H$) is the bandwidth of the deuterated
(hydrogenated) system and $g_{ihD}$ ($g_{ihH}$) is the coupling
between the $i^{th}$ mode and the HOMO in the deuterated
(hydrogenated) molecule. As both the deuterated and hydrogenated
systems are very close to the Mott transition \cite{Taniguchi} we
suggest that polarons may play a significant role in driving the
Mott transition by dueteration.

In conclusion we have demonstrated that by considering the forces on
the nuclei due to the addition or subtraction of an arbitrarily small
electronic charge
one may calculate the electron-phonon coupling constants as an ${\cal
O}(1)$ problem. This method is exact within the density functional
formalism and was shown to be numerically accurate for a large
number of small molecules. Note that, although we have only
considered molecular systems in our numerical work this is no
intrinsic limitation of this method which prevents it being applied
to infinite systems.

This work was motivated by conversations with Greg Freebairn. We
would like to thank James Annett, Stephen Dougdale, Nikitas
Gidopoulos, Barbara Montanari and Keith Refson for useful
conversations. This work was funded in part by the Australian Research
Council.  TB was supported by NSF NIRT-0304122. MRP was supported
in part by the HPCMO CHSSI program.
Some of the calculations were calculations were
performed on the Australian Partnership for Advanced Computing
(APAC) National Facility under a grant from the Merit Allocation
Scheme. Some of the calculations were performed on the HPCMO
computational platforms.


\begin{thebibliography}{99}



\bibitem{Ziman}
See for example, J.M. Ziman, `{\it Electrons and phonons}' (Oxford
University Press, Oxford, 1960).

\bibitem{Gruner}
For a recent review see, G. Gr{\"u}ner, `{\it Density waves in
solids}' (Perseus Publishing, Cambridge, 1994).

\bibitem{Mott&Alexandrov}
See, for example, A.S. Alexandrov and N.F. Mott, `{\it Polarons and
biploarons}' (World Scientific, Singapore, 1995).

\bibitem{Marcus}
For a review see, R.A. Marcus, Rev. Mod. Phys. 65, 599 (1993).

\bibitem{Bersuker}
See, for example, I.B. Bersuker, `{\it The Jahn-Teller effect and
vibronic interactions in modern chemistry}' (Plenum, New York,
1984).

\bibitem{Meredith}
P. Meredith and J. Riesz, Photochem. Photobiol. {\bf79}, 211 (2004).

\bibitem{Reimers}
J.R. Reimers and N.S. Hush, J. Am. Chem. Soc. {\bf126}, 4132
(2004).

\bibitem{rhodopsin}
S. Hahn and G. Stock, J. Phys. Chem. B {\bf104}, 1146 (2000).

\bibitem{DFPT}
%X. Gonze, Phys. Rev. A {\bf 52}, 1096 (1995);
S. Baroni, S. de
Gironcoli, A. D. Corso, and P. Giannozzi, Rev. Mod. Phys. {\bf 73},
515 (2001).

\bibitem{Janak}
J.F. Janak, Phys. Rev. B {\bf18}, 7165 (1978).

\bibitem{NRLMOL1}
M.R. Pederson and K.A. Jackson, Phys. Rev. B {\bf41}, 7453 (1990).

\bibitem{NRLMOL2}
K.A. Jackson and M.R. Pederson, Phys. Rev. B {\bf42}, 3276 (1990).

\bibitem{NRLMOL3}
M.R. Pederson and K.A. Jackson, Phys. Rev. B {\bf43}, 7312 (1991).

\bibitem{NRLMOL4}
A.A. Quong, M.R. Pederson and J.L. Feldman, Solid State Commun.
{\bf87}, 535 (1993).

\bibitem{NRLMOL5}
D.V. Porezag and M.R. Pederson, Phys. Rev. B {\bf54}, 7830 (1996).

\bibitem{NRLMOL6}
D.V. Porezag, Ph.D. Thesis, Technische Universitt, 1997, Available
from: http://archiv.tu-chemnitz.de/pub/1997/0025.

\bibitem{NRLMOL7}
A. Briley, M.R.
Pederson, K.A. Jackson, D.C. Patton and D.V. Porezag, Phys. Rev. B
{\bf58}, 1786 (1998).

\bibitem{PBE}
J.P. Perdew, K. Burke and M. Ernzerhof, Phys. Rev. Lett. {\bf77},
3865 (1996).

\bibitem{Taniguchi}
H. Taniguchi, K. Kanoda and A. Kawamoto, Phys. Rev. B {\bf67},
014510 (2003).

\end{thebibliography}
\end{document}